# Coexistence of ferromagnetism and ferroelectricity in the van der Waals multiferroic CuIn$_{0.2}$V$_{0.8}$P$_2$S$_6$

Subrata Ghosh[1,2,+,*], Rosalin Mohanty[3+], Yuwei Sun[3+], Soumi Mondal[2], Chandan De[2], Jose G. Jimenez[4], Weiwei Xie[4], Cheng Gong[3,*], and Zhiqiang Mao[1,2,5,*]

[1]*2D crystal consortium, Materials Research Institute, Pennsylvania State University, University Park, PA 16802, USA.*
[2]*Department of Physics, Pennsylvania State University, University Park, PA 16802, USA.*
[3]*Department of Electrical and Computer Engineering and Quantum Technology Center, University of Maryland, College Park, MD 20742, USA*
[4]*Department of Chemistry, Michigan State University, East Lansing, MI 48824, USA.*
[5]*Department of Materials Science and Engineering, Pennsylvania State University, University Park, PA 16802, USA*

*Corresponding authors' email*: smg7204@psu.edu (SG); gongc@umd.edu (CG); zim1@psu.edu (ZM)

## Abstract

Two-dimensional (2D) van der Waals (vdW) multiferroics have emerged as a promising platform for next-generation multifunctional devices. Although recent studies have demonstrated that artificial heterostructures can combine dual ferroic orders and exhibit strong magnetoelectric coupling, their performance is sometimes limited by poor interface quality and inadequate long-term stability. By contrast, the realization of intrinsic single-phase materials with coexisting ferromagnetism and ferroelectricity remains a longstanding challenge in the field. Here we report the realization of a single-phase 2D vdW multiferroic system, CuIn$_{0.2}$V$_{0.8}$P$_2$S$_6$, which exhibits both ferromagnetism and room-temperature ferroelectricity. The intrinsic ferroelectric nature of CuIn$_{0.2}$V$_{0.8}$P$_2$S$_6$ was probed using ferroelectric tunnel junctions, which exhibit a large tunneling electroresistance with an ON/OFF ratio of $10^7$ at 295 K. CuIn$_{0.2}$V$_{0.8}$P$_2$S$_6$ develops ferromagnetic ordering with the Curie temperature ($T_C$) of 14.6 K, as evidenced by pronounced magnetic hysteresis and a relatively large remanent magnetization. Notably, the appearance of a magnetodielectric response below $T_C$ is consistent with the anticipated interplay between the ferromagnetic and ferroelectric orders. These results highlight a promising route toward single-phase van der Waals multiferroics with coexisting ferroic orders.

[+]These authors equally contribute to this work

# Introduction

Ferroelectric materials, characterized by a spontaneous electric polarization that can be switched by an external electric field, have attracted significant interest for applications in non-volatile memory, logic devices, and sensors[1–6]. Multiferroics, which intrinsically combine ferroelectric and magnetic orders through magnetoelectric (ME) coupling, enable electric-field control of magnetism and vice versa, making them highly attractive for multifunctional device architectures[7,8]. However, maintaining robust ferroelectricity at room temperature in conventional ultrathin oxide multiferroics remains challenging due to the critical depolarization thickness, which imposes fundamental limitations on their scalability for nanoscale devices[9,10]. In this context, the emergence of two-dimensional (2D) van der Waals (vdW) ferroelectrics provides a promising pathway to overcome these constraints[11–17]. Their dangling-bond-free surfaces, weak interlayer vdW interactions, stable remanent polarization, and compatibility with atomic-scale integration make them particularly attractive for next-generation functional electronic devices[18].

The inherent conflict between long-range magnetic ordering and spontaneous electric polarization renders the realization of intrinsic single-phase 2D vdW multiferroics rare. In this context, the intrinsic 2D vdW system, such as $CuCrP_2S_6$[14,19], which shows the coexistence of antiferromagnetism (AFM) and antiferroelectricity (AFE), and $NiI_2$[20,21], where helical magnetic order drives FE, attract attention as promising multiferroics; however, their relatively weak macroscopic ME coupling hinders their practical applications. More recently, Tian et. al. reported an air-stable bilayer $CrTe_2$ that exhibits intrinsic room-temperature multiferroicity[22]. Alternatively, the dual ferroic orders, ferromagnetism (FM) and ferroelectricity (FE), are engineered within vdW multiferroic heterostructures through interfacial design strategies that facilitate strong ME coupling[23–28]. For instance, Liang et al. observed non-volatile electrical control of magnetism in 2D $Cr_2Ge_2Te_6$/FE polymer poly (vinylidene fluoride-co-trifluoroethylene) and $Fe_3GeTe_2/CuCrP_2S_6$ multiferroic heterostructures, demonstrating that the magnetic hysteresis loop could be effectively modulated via applying a small voltage due to strong interfacial ME coupling[25,28]. Similarly, Eom et al. showed the gate voltage control of magnetism in $Fe_{3-x}GeTe_2/In_2Se_3$ heterostructure owing to the presence of in-plane tensile strain[27]. Furthermore, Zhao et al. also successfully demonstrated non-volatile electrical control of FM in $Fe_3GaTe_2/CuInP_2S_6$ multiferroic heterostructures driven by FE polarization-enhanced Dzyaloshinskii-Moriya interaction (DMI) in FM[26]. While these artificial heterostructures can integrate ferromagnetism (FM) and ferroelectricity (FE) and exhibit magnetoelectric (ME) coupling through interfacial effects, their performance is sometimes limited by poor interface quality and inadequate long-term stability[1,22,29,30]. In contrast, the realization of single-phase two-dimensional multiferroics with coexisting ferroic orders could not only strengthen ME coupling through robust remanent magnetization and electric polarization but also help overcome limitations associated with engineered interfaces.

Here, we report the coexistence of ferromagnetism and ferroelectricity in a layered vdW $CuIn_{0.2}V_{0.8}P_2S_6$ system with ME coupling. The discovery of intrinsic room-temperature FE in ultrathin (~ 4 nm) flakes of $CuInP_2S_6$ (CIPS)[16] has stimulated extensive research on investigating ferroic order in Cu based-metal thiophosphates (MTPs) with the general formula $CuMP_2X_6$ (where M = In, Cr, V; X = S, Se), a vdW material family. In CIPS, FE polarization switching originates from the multiple occupations and migration dynamics of $Cu^+$ ions[31,32]; however, it is nonmagnetic. Recently, $CuVP_2S_6$ (CVPS) has been identified as an intriguing member, demonstrating intrinsic room-temperature FE while exhibiting FM-like behavior below Curie temperature ($T_C$) ~ 3.3 K with negligible magnetic hysteresis[33,34]. However, AC susceptibility measurement reveals a weak spin glass feature[34]. The emergence of diverse magnetic ground states within this material family highlights their potential as a versatile vdW platform for exploring multiferroicity.

Through systematic alloying of In and V at the In site in CIPS, we realize long-range ferromagnetic ordering in bulk $CuIn_{0.2}V_{0.8}P_2S_6$ single-crystals with a Curie temperature ($T_C$) of 14.6 K, while preserving room-temperature ferroelectricity. The intrinsic FM of $CuIn_{0.2}V_{0.8}P_2S_6$ is evidenced by the pronounced magnetic hysteresis with a relatively large remanent magnetization ($M_r$ ~ 0.06 $\mu_B$/f. u.), and coercivity ($H_C$) ~ 180 Oe at 2 K. Its intrinsic ferroelectric nature is confirmed by ferroelectric tunnel junctions, which display a large tunneling electroresistance with an ON/OFF ratio of $10^7$ at room temperature. In addition, the emergence of a magnetodielectric response below $T_c$ is consistent with the expected magnetoelectric coupling between the ferromagnetic and ferroelectric orders. Together, these results establish a promising route toward single-phase two-dimensional van der Waals multiferroics with coexisting ferromagnetism and ferroelectricity.

**Results and Discussions:**

**Synthesis and structural characterization of $CuIn_{1-x}V_xP_2S_6$ ($x$ = 0.05-1.0)**

CIPS is a layered vdW material of the MTP family, notable for its robust room temperature ferroelectricity[16]. It crystallizes in a monocline structure (space group: $Cc$) (**Fig. 1A**). The crystal structure of CIPS consists of a sulfur framework where the octahedral voids are occupied by the metal cations ($In^{3+}$, mobile $Cu^+$) and P-P pairs[16]. Its ferroelectricity arises along the out-of-plane direction due to the migration of $Cu^+$ ions to three distinct crystallographic sites under an external field[31]. CVPS adopts a similar crystal structure to CIPS[15]. We grew single crystals of $CuIn_{1-x}V_xP_2S_6$ (CIVPS) with $x$ = 0.05, 0.1, 0.2, 0.4, 0.8, and 1.0 using chemical vapor transport (CVT) (see Methods), and optical images of the as-grown single crystals are shown in **Fig. 1B**. To evaluate the crystal structure evolution in $CuIn_{1-x}V_xP_2S_6$, X-ray diffraction (XRD) measurement was performed on cleavage planes for all synthesised samples. The sharp (00$l$) diffraction peaks illustrated in **Fig. 1(B)** demonstrate good crystallinity of the CIVPS samples. Our single-crystal XRD refinements performed on CIVPS with $x$ = 0.8 and 1.0 indicate they also adopt the monoclinic $Cc$ structure, consistent with previous studies[15,16]. As shown in **Fig. 1B**, with increasing V-doping concentration, the (00$l$) peaks shift to higher angles, reflecting a reduction in the lattice constant along the $c$-axis and, consequently, a gradual decrease in the interlayer spacing. The actual compositions of the as-grown single crystals are determined by energy-dispersive X-ray spectroscopy (EDS), which varies from the nominal composition of the source material. The sample compositions presented below are the measured compositions (details are tabulated in Supplemental **Table S1**). Furthermore, the area mapping by EDS confirms the homogeneous distribution of constituent elements on a micrometre length scale for all the samples. **Figure 1C** shows a representative example for $CuIn_{0.2}V_{0.8}P_2S_6$.

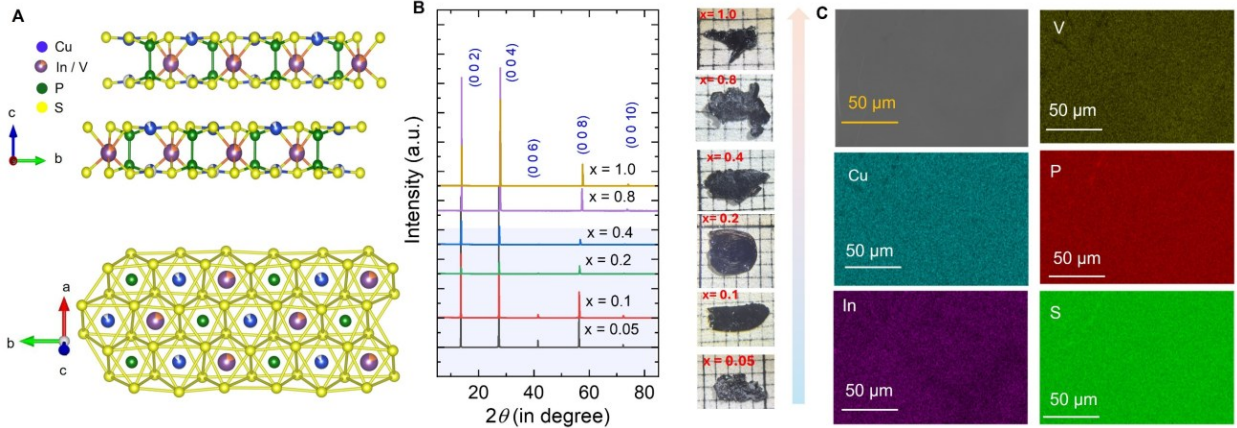

**Fig. 1| Structural and composition of CuIn$_{1-x}$V$_x$P$_2$S$_6$ ($x$ = 0.05-1.0).** (A) The side view and top view of the crystal structure of Cu(In,V)P$_2$S$_6$. (B) Room temperature XRD pattern of CuIn$_{1-x}$V$_x$P$_2$S$_6$ single crystals with their optical images. (C) EDS mapping in CuIn$_{0.2}$V$_{0.8}$P$_2$S$_6$ showing the homogeneous elemental distribution.

**Evolution of magnetism in CuIn$_{1-x}$V$_x$P$_2$S$_6$ ($x$ = 0.05-1.0)**

While CIPS is nonmagnetic, since neither Cu$^+$ nor In$^{3+}$ possesses partially filled $d$ orbitals, replacing In$^{3+}$ with magnetic cations such as Cr$^{3+}$ or V$^{3+}$ gives rise to diverse magnetic ground states, including AFM ordering in the Cr-substituted compound [14] and FM-like behavior in the V-substituted compound[33]. As shown below, we demonstrate that such tunability provides an effective means to stabilize long-range FM ordering through alloying of In$^{3+}$ and V$^{3+}$. Furthermore, the presence of heavier In$^{3+}$ relative to V$^{3+}$ enhances the spin-orbit coupling, thereby increasing magnetic anisotropy. To elucidate the magnetic properties of CuIn$_{1-x}$V$_x$P$_2$S$_6$ ($x$ = 0.05, 0.1, 0.2, 0.4, 0.8, 1.0), we measured the temperature dependence of zero-field-cooled (ZFC) and field-cooled (FC) magnetization under a magnetic field of ~ 0.1 T applied parallel to the $c$ axis ($H$ // $c$), as shown in **Fig. 2**. The low temperature magnetic susceptibility ($\chi_{dc}$) increases progressively with V-doping and reaches a maximum at $x$ = 0.8 (**Fig. 2G**). Upon further increasing the V content to $x$ = 1.0 (CVPS, $\chi_{dc}$ decreases relative to the $x$ = 0.8 composition.

Moreover, we also measured the isothermal magnetization ($M$-$H$) curve of CuIn$_{1-x}$V$_x$P$_2$S$_6$ at 2 K under $H$ // c, as shown in the **insets of Fig. 2(A-F)**. The polarized magnetization ($M_P$) of CuIn$_{1-x}$V$_x$P$_2$S$_6$ increases systematically with V content; under a 7 T magnetic field, $M_P$ reaches approximately 0.12, 0.17, 0.37, 0.94, 1.14, 1.45 μ$_B$/f. u. for $x$ = 0.05, 0.1, 0.2, 0.4, 0.8, 1.0, respectively (**Fig. 2I**). The combined data of susceptibility and isothermal $M$-$H$ measurement suggest that the $x$ = 0.05-0.4 samples exhibit paramagnetic behavior with short-range FM correlations emerging at low temperature, as evidenced by the broad FM-like polarization under magnetic field (**insets of Fig. 2A- 2D**). The $x$ = 1.0 sample shows FM-like behaviour with a weak spin-glass state as evident from the magnetic relaxation measurements (see **Fig. S1, Supplementary Information**), consistent with the prior report[34]. In contrast, CuIn$_{0.2}$V$_{0.8}$P$_2$S$_6$ (CIVPS, $x$ = 0.8) demonstrates a typical long-range FM behaviour as evidenced by a sharp transition in $\chi_{dc}$ (**Fig. 2E**), step-like FM polarization under field (**inset of Fig. 2E**), pronounced magnetic hysteresis (as discussed later), and the absence of any spin-glass state in magnetic relaxation measurements (see **Fig. S1**, Supplementary Information). Further discussions about its intrinsic ferromagnetic nature are presented in the next section.

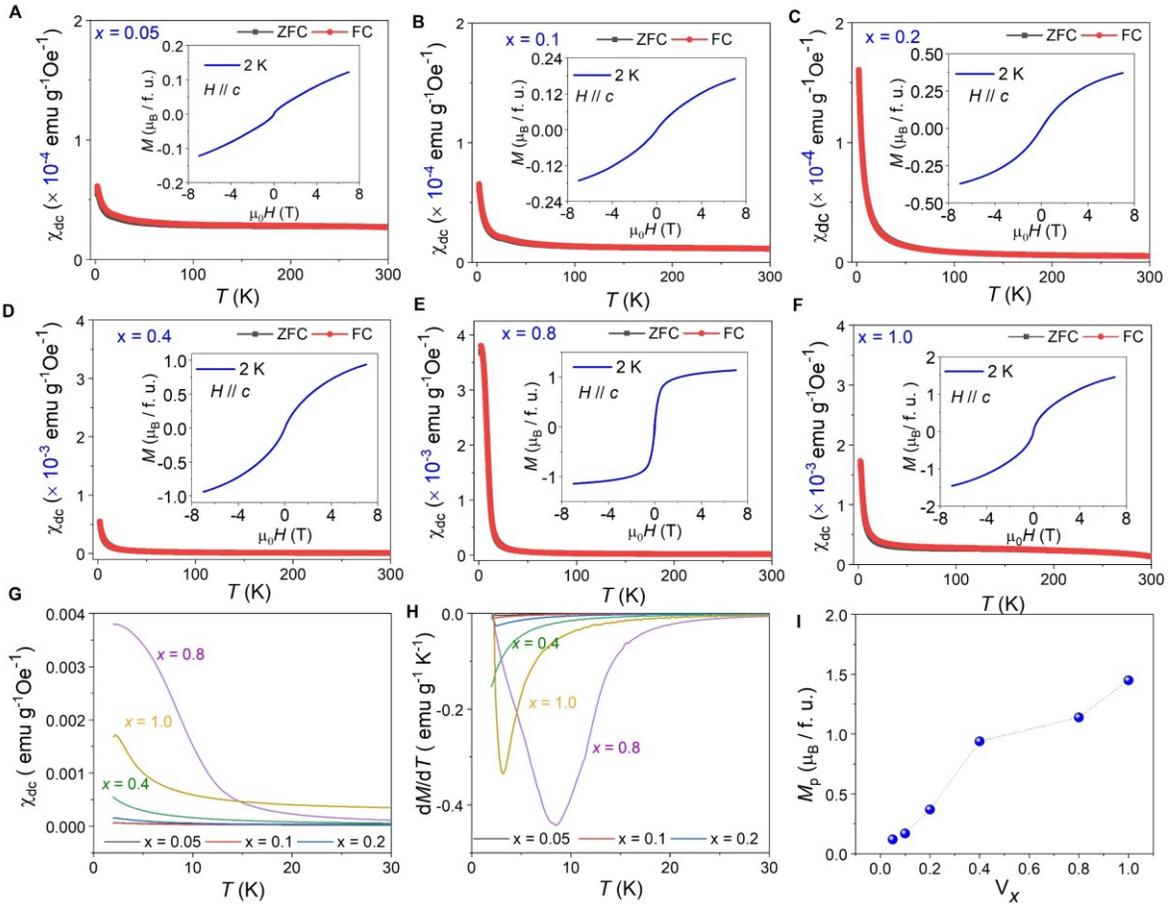

**Fig. 2| Magnetic properties of CuIn$_{1-x}$V$_x$P$_2$S$_6$. (A-F)** Temperature dependence of dc magnetic susceptibility of (A) $x = 0.05$, (B) $x = 0.1$, (C) $x = 0.2$, (D) $x = 0.4$, (E) $x = 0.8$, and (F) $x = 1.0$ in zero-field-cooled (ZFC) and field-cooled (FC) conditions under a magnetic field of 0.1T, applied out of plane direction ($H // c$). Note the panels (A-C) share a common y-axis scale (factor of $10^{-4}$) while panels (D-F) are plotted on the same scale with a factor of $10^{-3}$. Inset: Corresponding isothermal *M-H* curve at 2 K under $H // c$. (G) FC $\chi_{dc}$ of all the samples are plotted in the same panel at low temperature regime for comparison. (H) dM/dT vs. T plot reveals the magnetic ordering temperature. (I) Under a 7T magnetic field, the $M_p$ value is plotted with V doping concentration.

### Intrinsic ferromagnetism and magnetic anisotropy in CuIn$_{0.2}$V$_{0.8}$P$_2$S$_6$ (CIVPS, $x = 0.8$)

The long-range FM state in CIVPS ($x = 0.8$) is further illustrated by the temperature dependent $\chi_{dc}$ measurement performed at a low applied field of ~ 0.03T (note that the measurement shown in Fig. 2 was done under 0.1 T), under ZFC and FC conditions, with the magnetic field applied along the out-of-plane ($H // c$) and in-plane ($H // ab$) directions (**Fig. 3A**). The paramagnetic-to-FM transition temperature was determined from the minimum in d*M*/d*T* vs *T* curve, yielding a Curie temperature ($T_C$) of 8.5 K, with an onset near 14.6 K (**Fig. 2H**). A pronounced bifurcation between ZFC-FC magnetization curves appears below $T_C$, which is more prominent for *H // c* and not associated with a glassy state, as mentioned earlier. This behavior suggests a strong magnetocrystalline anisotropy, as further evidenced by the isothermal *M-H* measurement. **Figure 3B** presents isothermal *M-H* data at 2 K under *H // c* and *H // ab* conditions. Although $\chi_{dc}$ is higher for *H // ab* than for *H // c* (**Fig. 3A**), the saturation magnetization is 35 % larger for *H // c*, accompanied by a pronounced magnetic hysteresis (**Fig. 3C**) that appears to suggest the *c*-axis is the spin easy axis. However, the critical field required to reach saturation is 25% lower for *H // ab* than for *H // c*, contradicting a simple *c*-axis easy-axis picture. These observations instead suggest that the spin easy axis is not exactly along the *c*-axis but is likely tilted and close to it. Similar unusual magnetic anisotropy has previously been reported in the vdW ferromagnet VI$_3$[35,36]. A precise determination of the spin orientation, however, would require neutron scattering and is beyond the scope of this work.

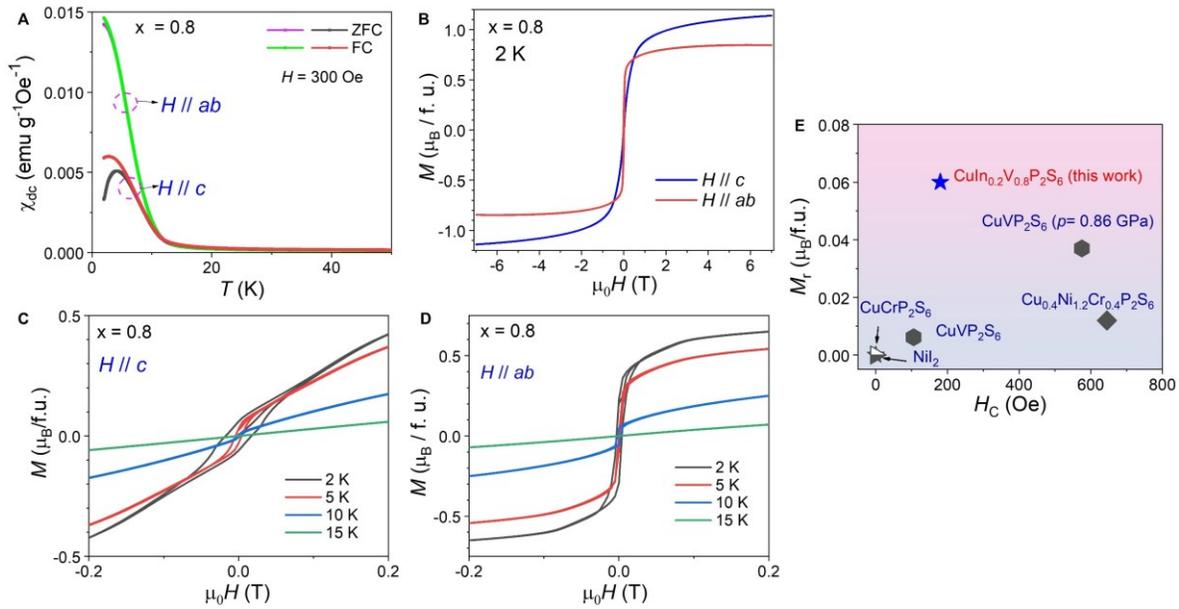

**Fig. 3| Ferromagnetism in CuIn$_{0.2}$V$_{0.8}$P$_2$S$_6$.** (A) The temperature dependence of *dc* magnetic susceptibility under ZFC and FC conditions for both *H //c* and *H // ab*. (B) Isothermal *M-H* curves at 2 K under *H //c* and *H // ab*. (C, D) Isothermal *M-H* curves at various temperatures within the low field regime for (C) *H //c* and (D) *H // ab*. (E) The M$_r$ and H$_c$ of CuIn$_{0.2}$V$_{0.8}$P$_2$S$_6$ are compared with reported vdW multiferroics[19,33,37,38].

In contrast, the *x* = 1.0 (CVPS) sample exhibits FM-like behavior, manifested by broad FM polarization (**Fig. 2F**). $T_C$ of CVPS is 3.3 K, determined from the d*M*/d*T* vs *T* plot, with an onset near 6.5 K (**Fig. 2H**), which is lower compared to the *x* = 0.8 sample. CVPS also shows a spin glass state, as evidenced by magnetic relaxation measurements, as discussed earlier. Isothermal *M-H* measurements at 2 K under both *H // c* and H*//ab* conditions (see **Fig. S2A,** supplementary information) reveals while it shows different polarized magnetization, around 18 % larger for *H // c* compared to *H // b*, it shows relatively weak magnetic anisotropy compared to *x* = 0.8. Notably, we did not observe magnetic hysteresis for CVPS **(see Fig. S2B,** supplementary information**)**, in agreement with prior studies[34]. The comparison between *x* = 0.8 and 1.0 highlights that alloying stabilizes the long-range FM order and significantly enhances magnetic anisotropy in CIVPS with *x* = 0.8. The enhanced magnetic anisotropy is attributed to increased spin-orbit coupling arising from the incorporation of heavier In atoms. The large magnetocrystalline anisotropy highlights CIVPS($x = 0.8$) as a promising van der Waals ferromagnet.

Further insight is provided by low-field ($\pm 0.2$ T) isothermal $M$-$H$ measurements at various temperatures (**Fig. 3C and Fig. 3D**), which clearly reveal the magnetic hysteresis in CIVPS ($x = 0.8$) below $T_C$ for both $H // c$ and $H // ab$. The hysteresis, however, is substantially larger for $H // c$, with a relatively large remanent magnetization ($M_r$) of ~ 0.06 $\mu_B$/f. u. and a coercive field ($H_C$) of ~180 Oe at 2 K. The presence of magnetic hysteresis unambiguously verifies the FM ordering in CIVPS ($x = 0.8$). Notably, the observed $M_r$ value is significantly larger than those reported for previously known vdW multiferroics (**Fig. 3E**)[19,33,37,38], underscoring the potential of CIVPS ($x = 0.8$) as a vdW multiferroic with enhanced ME coupling.

**Intrinsic room-temperature ferroelectricity in CuIn$_{0.2}$V$_{0.8}$P$_2$S$_6$ (CIVPS, $x = 0.8$)**

To probe the ferroelectricity of CIVPS ($x = 0.8$), ferroelectric tunnel junctions (FTJs), consisting of a thin FE layer between two electrodes, offer an effective approach via the tunneling electroresistance (TER) effect[39–41]. In such devices, the FE layer serves as a quantum tunneling barrier, where reversal of spontaneous polarization modulates the tunneling conductance by altering the interfacial band alignment, resulting in distinct low- (ON) and high-resistance (OFF) states. The intrinsic ferroelectricity of CIVPS ($x = 0.8$) was investigated by fabricating FTJs using mechanically exfoliated CIVPS ($x = 0.8$) flakes on a 260-nm-thick SiO$_2$/Si substrate. The device structure (**Fig. 4A**) consists of a 30 nm Au (metal) bottom electrode, a thin CIVPS ($x = 0.8$) flake as the FE spacer, and a few-layer graphene (semimetal) top electrode (see methods). The thicknesses of CIVPS and graphene were measured by atomic force microscopy to be approximately 12 nm and 9 nm, respectively, as shown in **Fig. 4B**. To probe FE switching, voltage pulses of opposite polarities were applied to the junction to switch the FE polarization between the two electrodes and thereby modulate the resistance state. A negative voltage pulse (−8 V) aligned the FE polarization toward the Au electrode, defining the ON state, while a positive pulse (+8 V) switched the polarization toward the graphene electrode, corresponding to the OFF state (**see inset, Fig. 4C**). Here, the ON and OFF labels are assigned based on the measured low and high tunneling current densities, respectively, as shown in **Fig. 4C**. The device exhibits large ON/OFF ratios on the order of $10^7$ at room temperature (295 K), evaluated at a reading voltage of 0.2 V. Such a resistance switch with a large ON/OFF ratio provides strong evidence of polarization-controlled transport in CIVPS ($x = 0.8$)-based FTJs. Furthermore, temperature-dependent current density-voltage ($J$-$V$) measurements (**Fig. 4D**) in the ON state exhibit increasingly nonlinear behavior with decreasing temperature, indicating the presence of finite Schottky barriers at the Au/CIVPS ($x = 0.8$) and graphene/CIVPS ($x = 0.8$) interfaces[40,42]. These results confirm that transport is dominated by polarization-modulated tunneling, demonstrating the FE nature of the material.

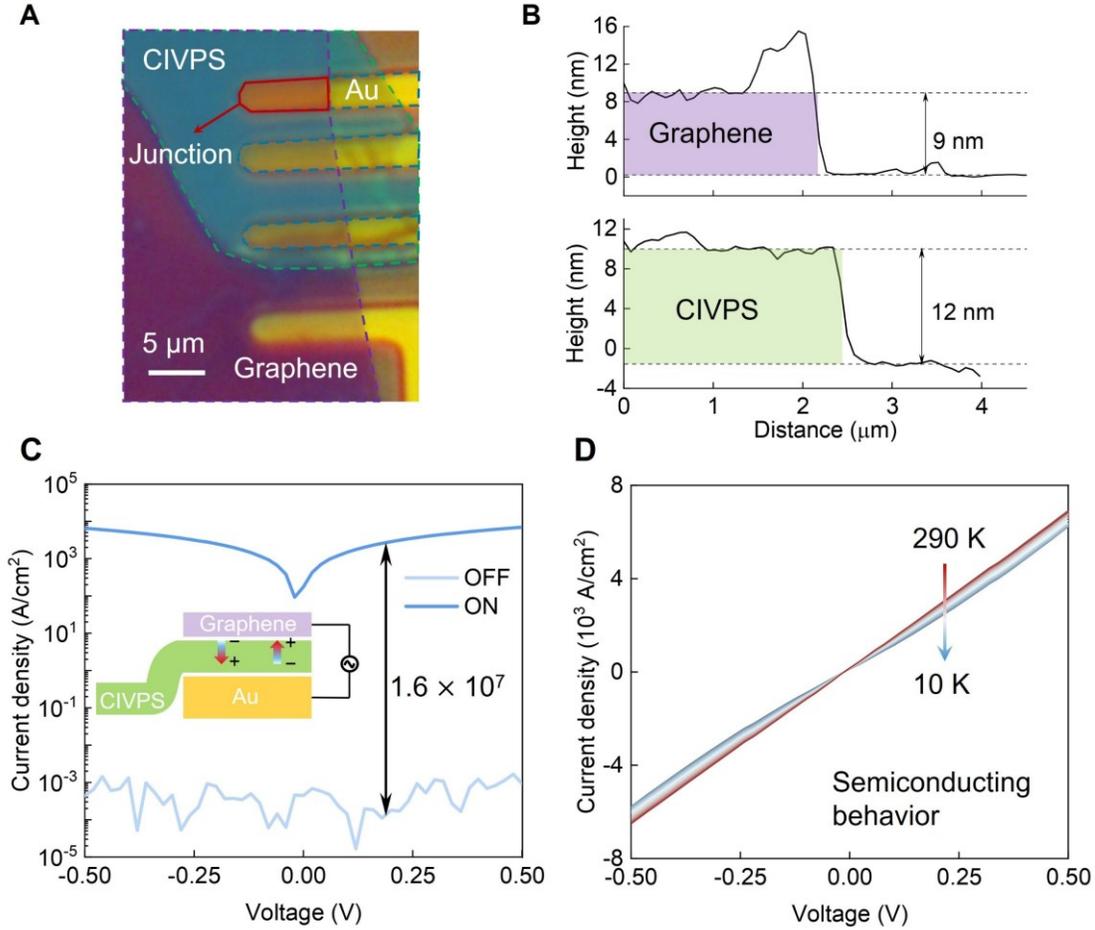

**Fig. 4 | Ferroelectricity in CIVPS ($x = 0.8$) flakes.** (**A**), An optical image of an FTJ device composed of a graphene top electrode, a CIVPS ($x = 0.8$) FE spacer, and an Au bottom electrode on a 260-nm SiO$_2$/Si chip. (**B**), Corresponding atomic force microscopy height profiles of the graphene and CIVPS flakes. The atomic force microscopy measurements indicate a thickness of 9 nm for the top graphene electrode and 12 nm for the CIVPS ($x = 0.8$) spacer. (**C**), Current density versus reading voltage for the ON and OFF states of the FTJ device, measured from −0.5 V to +0.5 V at room temperature (295 K). The ON/OFF ratios (probed by 0.2 V reading voltage) were measured to be $1.6 \times 10^7$. The inset illustrates the device schematic. (**D**), The temperature-dependent ON-state current density-voltage (*J-V*) characteristics measured from 290 K to 10 K for the FTJ.

**Coupling of ferromagnetism and ferroelectricity in CuIn$_{0.2}$V$_{0.8}$P$_2$S$_6$ (CIVPS, $x = 0.8$)**

      Our results discussed above explicitly prove the coexistence of intrinsic ferromagnetism and ferroelectricity in CIVPS ($x = 0.8$). Given the anticipated coupling between magnetic order and electrical polarization, we further investigated the magnetodielectric properties of the CIVPS ($x = 0.8$) sample. We measured the temperature dependence of the dielectric constant ($\varepsilon_r$) along the *c* axis and the magnetic field-dependent $\varepsilon_r$ at different temperatures. The temperature-dependent $\varepsilon_r$ within the frequency range of 1-100 kHz, as illustrated in **Fig. 5A**, reveals a pronounced anomaly coinciding with the onset of $T_C$ (~14.6 K), consistent with the expected magnetoelectric coupling in CIVPS ($x = 0.8$). Although $\varepsilon_r$ reduces slightly with increasing frequency, as expected, the anomaly in $\varepsilon_r$ remains invariant with increasing frequency, indicating the intrinsic coupling to the FM ordering[43]. The coupling between magnetic order and electrical polarization in CIVPS

indicates the strong spin-phonon coupling that can arise from the exchange striction mechanism or through the orbital degree of freedom[44].

To further elucidate the ME coupling, we measured the magnetic field dependence of $\varepsilon_r$ across a range of temperatures. As shown in **Fig. 5B**, CIVPS ($x = 0.8$) shows a striking magnetodielectric effect at 2 K. The normalized field-dependent $\varepsilon_r$, defined as $\varepsilon_r(H)/\varepsilon_r(H=0)$, is one order of magnitude higher than that of $CuCrP_2S_6$, a material from the same family, exhibiting a coupling between AFE and AFM states[34]. Furthermore, the magnetodielectric response diminishes as the temperature approaches $T_C$ and disappears completely above $T_C$. This temperature-dependent behavior further implies an interplay between the ferroelectric and ferromagnetic orders in CIVPS ($x = 0.8$).

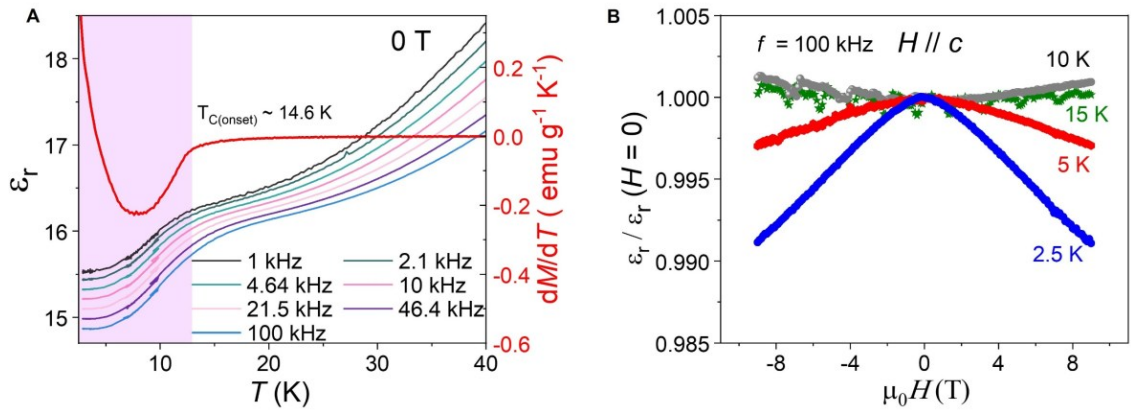

**Fig. 5| Magnetodielectric properties in CIVPS ($x = 0.8$).** (A) Temperature dependence of dielectric constant ($\varepsilon_r$) measured along the out-of-plane direction during cooling, at various frequencies, ranging from 1 kHz to 100 kHz. The $dM/dT$ vs $T$ under $H // c$ is plotted in the same graph (solid blue line, right Y axis) to demonstrate the anomaly in $\varepsilon_r$ at Tc (B) Normalized field-dependent $\varepsilon_r$ at different temperatures at $f = $ 100 kHz. The magnetic field was applied along the c-axis.

**Conclusion**

In summary, we have successfully synthesized a single-phase vdW multiferroic $CuIn_{0.2}V_{0.8}P_2S_6$, through systematic alloying of V and In, and demonstrated the coexistence of intrinsic ferromagnetism and ferroelectricity. The ferroelectric nature of $CuIn_{0.2}V_{0.8}P_2S_6$ was confirmed by the fabrication of ferroelectric tunnel junctions, which exhibits a remarkable tunnelling electroresistance effect with an ON/OFF ratio of $10^7$ at room temperature. In addition, $CuIn_{0.2}V_{0.8}P_2S_6$ develops intrinsic ferromagnetic ordering below 14.6 K, characterized by large remanent magnetization and strong magnetic anisotropy. The observation of the magnetodielectric effect below $T_C$ is consistent with the anticipated magnetoelectric coupling between the two ferroic orders. The realization of single-phase vdW multiferroicity in $CuIn_{0.2}V_{0.8}P_2S_6$ establishes it as a versatile platform for the development of next-generation multifunctional electronic and spintronic devices.

**Methods**

**Synthesis of CuIn$_{1-x}$V$_x$P$_2$S$_6$ single crystal**

Single crystals of CuIn$_{1-x}$V$_x$P$_2$S$_6$ ($x = 0.05$–1.0) were synthesized by chemical vapor transport (CVT). High-purity Cu, In, V, P, and S powders were loaded into quartz tubes under an inert atmosphere, using the molar ratios listed in Supplementary Table S1. The source materials were thoroughly mixed to ensure compositional homogeneity, and the quartz tube was sealed under vacuum (~ $10^{-5}$ torr) after adding 50 mg of iodine as a transport agent. The sealed ampoule was then placed horizontally in a two-zone CVT furnace. Initially, the temperatures of the evaporation and crystallization zones were maintained at 650 °C and 750 °C for 2 days. After this, the evaporation and crystallization zones were set to 700 °C and 665 °C, respectively, and held for 5 days to promote crystal growth via vapor transport. Finally, the furnace was cooled down to room temperature over a period of 15 hours. The as-grown crystals appear as shiny black flakes with the typical longest dimension of 1-4 mm. Note that the nominal composition differs from the actual composition of the crystals (see Table S1).

**Structural and chemical composition analysis**

Room temperature X-ray diffraction patterns of our as-grown single crystals were performed using an X-ray diffractometer (Malvern PANalytical Empyrean III and IV). The actual composition and the compositional homogeneity of the single crystals were checked by energy dispersive X-ray spectroscopy (EDS, Oxford Aztec) analysis.

**Magnetic, dielectric, and magnetodielectric measurements**

Magnetic measurements were performed in a vibrating-sample SQUID magnetometer (Quantum Design, SQUID-VSM). Temperature and magnetic field-dependent capacitance were measured in a Physical Property Measurement System (PPMS) using an Agilent E4980A Precision LCR meter. The dielectric constant value was extracted from the measured capacitance value. For these measurements, Au was evaporated on both flat surfaces of the samples, and electrodes were made using silver-epoxy paste.

**FTJ device fabrication**

The Au electrodes were patterned using a Heidelberg MLA150 maskless aligner with S1813 photoresist, followed by development in CD-26 developer for 20 s. Au contacts of 30 nm thickness were then deposited in an Angstrom E-beam evaporator at a rate of 1 Å/s. The lift-off process was performed by immersing the samples in acetone for 1 hour, followed by rinsing in isopropyl alcohol (IPA) and drying under a compressed air flow. Atomically thin CuIn$_{0.2}$V$_{0.8}$P$_2$S$_6$ flakes were mechanically exfoliated from their bulk crystals onto polydimethylsiloxane (PDMS) stamps via adhesive tape, and their thicknesses were preliminarily identified by optical contrast under a microscope. The selected CuIn$_{0.2}$V$_{0.8}$P$_2$S$_6$ flakes were then transferred onto the pre-patterned Au electrodes using an all-dry viscoelastic transfer technique. Finally, a graphene top electrode was transferred using the same PDMS stamping method to complete the device fabrication.

**Electrical characterization**

The *J-V* measurements were performed using a Keithley 2450 source-measure unit. Electrical transport measurements were performed in a Montana cryostat with a base temperature of 4 K. Samples were mounted onto a twelve-line sample holder using conductive silver paste, followed by 50 nm Au wire bonding and loading into the cryostat. FE polarization was switched using voltage pulses to minimize Joule heating effects. The pulses were generated by an Agilent 81104A pulse/pattern generator with a pulse width of 1 ms and a repetition period of 1 s, and were applied for a total duration of 5 s. The pulse amplitude was increased in steps of 0.2 V until the CuIn$_{0.2}$V$_{0.8}$P$_2$S$_6$ layer was fully polarized (approximately 8 V), as indicated by the saturation of the *J-V* characteristics at higher pulse amplitudes. ON/OFF ratios were extracted at room temperature under a bias voltage of 0.2 V.

**Acknowledgement**

The material synthesis and characterization in this work were supported by the Pennsylvania State University Two-Dimensional Crystal Consortium–Materials Innovation Platform (2DCC-MIP), supported by the NSF under Cooperative Agreement No. DMR-2039351. Mao and Cheng also acknowledge support from the NSF under Grant No. DMR-2425599.

Supplementary Information

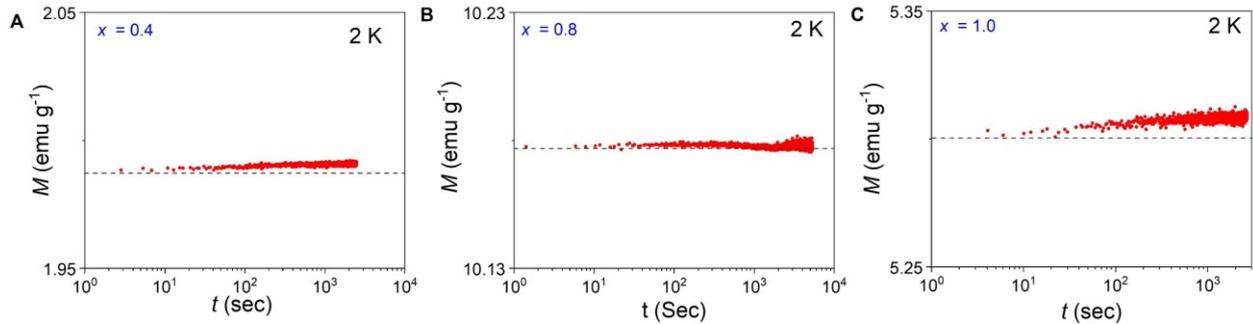

**Fig. S1| Magnetic relaxation measurement.** Magnetic relaxation measurement of $x$ = 0.4, 0.8, and 1.0 samples at 2K. The sample was cooled to 2 K, and a magnetic field of 0.5 T was applied; subsequently, the magnetization was measured as a function of time. The 'Y' axis width is similar for all three panels, and a solid dotted black straight line is added to guide the behavior of the data. Clearly, while the $x$ = 0.4 and 1.0 samples show a weak spin-glass feature, $x$ = 0.8 rules out the spin-glass state.

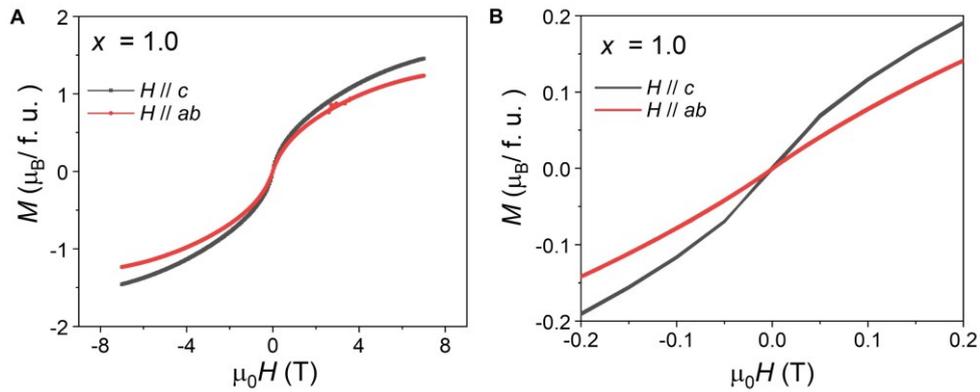

**Fig. S2| Magnetic properties of $x$ = 1.0 (CVPS).** (A) Isothermal $M$-$H$ curves at 2 K under $H$ //$c$ and $H$ // $ab$. (B) Isothermal $M$-$H$ curves at 2K within the low field regime for $H$ //$c$ and $H$ // $ab$.

**Table S1.** Nominal and actual composition of the $CuIn_{1-x}V_xP_2S_6$ ($x$ = 0.05-1.0) samples.

| Nominal Composition | Actual Composition from EDS | Termed as |
|---|---|---|
| $CuIn_{0.9}V_{0.1}P_2S_6$ | $Cu_{1.3}\mathbf{In_{0.97}V_{0.04}}P_{1.85}S_{5.85}$ | $x$ = 0.05 |
| $CuIn_{0.8}V_{0.2}P_2S_6$ | $Cu_{1.16}\mathbf{In_{0.96}V_{0.1}}P_{1.85}S_{5.93}$ | $x$ = 0.1 |
| $CuIn_{0.8}V_{0.4}P_2S_6$ | $Cu_{1.28}\mathbf{In_{0.79}V_{0.22}}P_{1.83}S_{5.88}$ | $x$ = 0.2 |
| $CuIn_{0.4}V_{0.6}P_2S_6$ | $Cu_{1.26}\mathbf{In_{0.64}V_{0.38}}P_{1.84}S_{5.88}$ | $x$ = 0.4 |
| $CuIn_{0.2}V_{0.8}P_2S_6$ | $Cu_{1.18}\mathbf{In_{0.23}V_{0.8}}P_{1.86}S_{5.93}$ | $x$ = 0.8 |
| $CuVP_2S_6$ | $Cu_{1.14}\mathbf{V_{1.06}}P_{1.83}S_{5.96}$ | $x$ = 1.0 |